\documentclass[preprint,10pt]{elsarticle}

\usepackage{graphicx}%
\usepackage{multirow}%
\usepackage{amsmath,amssymb,amsfonts}%
\usepackage{amsthm}%
\usepackage{mathrsfs}%
\usepackage[title]{appendix}%
\usepackage{xcolor, soul}%
\usepackage{textcomp}%
\usepackage{manyfoot}%
\usepackage{booktabs}%
\usepackage{algorithm}%
\usepackage{algorithmicx}%
\usepackage{algpseudocode}%
\usepackage{listings}%
\usepackage{verbatim}
\usepackage{subcaption}
\usepackage{graphicx}
\usepackage{physics}
\usepackage{url}
\usepackage{units}
\usepackage{gensymb}
\usepackage{ulem}
\biboptions{numbers,sort&compress}

\newcommand{\ra}[1]{\renewcommand{\arraystretch}{#1}}

\def\etal.{et\penalty50\ al.}
\graphicspath{ {figures/} }

\journal{Journal of Biomechanical Engineering}

\newcommand{\revs}[1]{{#1}}
\newcommand{\reva}[1]{{#1}}
\newcommand{\revb}[1]{{#1}}

\begin{document}

\begin{frontmatter}

\title{Influence of coronary plaque morphology on local mechanical states and associated in-stent restenosis}

\author[1,2]{Janina C. Datz\corref{cor1}}\ead{janina.datz@tum.de}
\author[3]{Ivo Steinbrecher}
\author[2]{Johannes Krefting}
\author[2]{Leif-Christopher Engel}
\author[3]{Alexander Popp}
\author[4]{Martin R. Pfaller}
\author[2]{Heribert Schunkert}
\author[1,5]{Wolfgang A. Wall}

\cortext[cor1]{Corresponding author}

\affiliation[1]{organization={Institute for Computational Mechanics, Technical University of Munich}, country={Germany}}
\affiliation[2]{organization={Department of Cardiology, TUM University Hospital German Heart Center, Technical University of Munich}, country={Germany}}
\affiliation[3]{organization={Institute for Mathematics and Computer-Based Simulation, University of the Bundeswehr Munich}, country={Germany}}
\affiliation[4]{organization={Biomedical Engineering, Yale University}, country={USA}}
\affiliation[5]{organization={Munich Institute of Biomedical Engineering, Technical University of Munich}, country={Germany}}

\begin{abstract}

In-stent restenosis \revb{(ISR)} after percutaneous coronary intervention is a multifactorial process. 
Specific morphological lesion characteristics were observed to contribute to the occurrence of \revb{ISR}.
Local mechanical factors, such as stresses and strains, are known to influence tissue adaptation after stent implantation. 
However, the influence of morphological features on those local mechanical states and, hence, on the occurrence of \revb{ISR} remains understudied. 
This work \revb{explores how} local mechanical quantities \revb{relate to ISR} by evaluating the stress distributions in the artery wall during and after stent implantation for \revb{morphology-informed lesion examples}.
We perform computational simulations of the stenting procedure with physics-based patient-specific coronary artery models. 
Different morphologies are assessed using the spatial plaque composition information from high-resolution coronary computed tomography angiography data.
\revb{In the sample cases, elevated local tensile stresses were consistently found at sites corresponding to ISR.}
We found that specific morphological characteristics like circumferential or asymmetric block calcifications result in higher stresses in the surrounding tissue. 
\revb{These findings show that for the observed connection between plaque morphology and ISR, the local mechanical state may represent a relevant link.}
\revb{This study provides a mechanistic, illustrative insight for the examined cases.}
\revb{Future work with larger cohorts and systematic follow-up can establish statistically robust associations.}

\end{abstract}



\end{frontmatter}


\section{Introduction}

Percutaneous coronary intervention (PCI) with stent implantation is the most frequently used treatment for flow-limiting plaques in coronary arteries. 
However, in up to 10\% of treated lesions, \revb{ISR} is diagnosed at follow-up after 6-12 months~\cite{byrneStentThrombosisRestenosis2015}.
\revb{ISR is defined as} a reocclusion of the vessel by more than 50\% of the initial lumen \revb{within the stented segment, i.e., the stent and \unit[5]{mm} adjacent to its borders~\cite{stefanini_ManagementMyocardialRevascularisation_}.}
The occurrence of ISR depends on a complex interplay between patient-, lesion-, and procedural-specific factors~\cite{casseseIncidencePredictorsRestenosis2014}.
A thorough assessment of a lesion's risk for stenting complications is crucial for choosing an appropriate lesion preparation technique, determining the optimal treatment strategy, and planning follow-up examinations. 
However, the known risk factors are insufficient to predict ISR reliably.
Importantly, individual physical properties of the lesion are currently not considered. 
Coronary imaging, such as optical coherence tomography (OCT) or intravascular ultrasound, has become increasingly important in PCI in assessing complex lesions and gathering more information about individual lesion characteristics. 
Observational clinical studies suggest that specific lesion morphologies influence the ISR risk~\cite{shafiabadihassani_InStentRestenosisOverview_2024, adolf_SpecificCalciumDeposition_2024, suzuki_ClinicalOutcomesPercutaneous_2024,dawson_HighRiskCoronaryPlaque_2022, iraqi_InfluencePlaqueCharacteristics_2025, tajima_AdvancedCTImaging_2024}.
Utilizing lesion morphology information could add a key element to the traditional risk assessment factors.

The currently known risk factors for ISR in clinical guidelines were determined in large-scale clinical studies. 
For instance, a small target vessel size or increased stented length were found to be independent predictors for ISR~\cite{casseseIncidencePredictorsRestenosis2014,guldener_MachineLearningIdentifies_2023}. 
While these studies found strong correlations between lesion-specific characteristics and the occurrence of ISR, local mechanical factors, i.e., continuum metrics such as stress, strain, stiffness, or internal elastic energy, were not considered.

During PCI, the artery tissue undergoes large stretches beyond its normal physiological domain, which can promote inflammatory pathways that serve as a stimulus for the formation of neointimal tissue and restenosis~\cite{lally_SimulationInstentRestenosis_2006, rodriguez-granillo_CoronaryArteryRemodelling_2006, liu_RiskFactorsInstent_2024}.
Additionally, it has been shown that changes in the mechanical state strongly affect vascular growth mechanisms, as cell and matrix turnover and growth mechanisms work towards restoring the tissue's natural mechanical homeostasis, i.e., the set point of the tissue's preferred mechanical state~\cite{ironsIntracellularSignalingControl2022,eichingerMechanicalHomeostasisTissue2021}.
The mechanical state is sensed by cells, and deviations from the homeostatic targets stimulate modifications in their phenotype and extracellular matrix~\cite{eichinger_WhatCellsRegulate_2021,humphreyVascularMechanobiologyHomeostasis2021}.
Hence, significant elevations of the stress values above a critical threshold outside the normal physiological range influence growth and remodeling processes.

Local material properties in diseased coronary arteries differ depending on the individual plaque composition. 
Lipid-rich tissue is a soft, fatty material; fibrous plaque tissue is generally stiffer, and calcified tissue behaves rigidly. 
It has been shown that the plaque composition and morphology influence the mechanical state during PCI~\cite{farbMorphologicalPredictorsRestenosis2002, brown_RoleBiomechanicalForces_2016}.
Furthermore, several studies suggest a connection between the clinical outcome and the plaque morphology. 
Morphological characteristics of the plaque include specific structural features, such as the location, shape, and size of different plaque components.  
For instance, circumferential or convex calcifications, eccentric plaque, and the ``headlight sign'', i.e., blocks of calcium on opposing sides of the vessel cross-section, have been observed to pose a high risk for stent underexpansion and ISR~\cite{shafiabadihassani_InStentRestenosisOverview_2024, adolf_SpecificCalciumDeposition_2024, suzuki_ClinicalOutcomesPercutaneous_2024}. 
Clinical studies on the role of advanced plaque features, including the coronary calcium extent, position, morphology, and density, were reviewed regarding the risk of acute coronary events~\cite{dawson_HighRiskCoronaryPlaque_2022}, ischaemia~\cite{iraqi_InfluencePlaqueCharacteristics_2025}, and PCI complications~\cite{tajima_AdvancedCTImaging_2024}.
The latter concluded that detailed plaque characterization is crucial for tailored and optimized PCI interventions.
However, the correlation between plaque morphological features, their influence on the tissue's mechanical state, and the associated clinical outcome remain underexplored.

Computational modeling is well established for studying structural mechanics in PCI.
Different approaches for finite element simulations of the stenting procedure in patient-specific coronary artery models have been proposed by~\cite{mortier_PatientspecificComputerModelling_2015,datz_PatientspecificCoronaryAngioplasty_2025,morlacchi_PatientspecificSimulationsStenting_2013,auricchio_PatientspecificFiniteElement_2013a,ragkousis_ComputationalModellingMultifolded_2015,ragkousis_SimulationLongitudinalStent_2014,mortier_NovelSimulationStrategy_2010,zahedmanesh_SimulationBalloonExpandable_2010,gijsen_SimulationStentDeployment_2008}.
However, information about the plaque morphology, which is critical for predicting stress distributions, was not included.

The impact of the overall plaque type, i.e., cellular, hypocellular, or (partially) calcified, on the lumen gain and resulting wall stresses during PCI was shown in generic artery models~\cite{conway_NumericalSimulationStent_2017,karimi_NonlinearFiniteElement_2014,pericevic_InfluencePlaqueComposition_2009,conway_ModellingAtheroscleroticPlaque_2014,helou_InfluenceBalloonDesign_2021}. 
Generic artery models with spatially distinguished plaque materials were used in~\cite{deokar_ComputationalModelingComparative_2017,wei_InfluencesPlaqueEccentricity_2019,morlacchi_InfluencePlaqueCalcifications_2014,iannaccone_InfluenceVascularAnatomy_2014}, demonstrating the effect of different plaque morphologies.
However, the characteristics of realistic arteries differ significantly from generic plaque models. 
Recent studies used 2D and 3D patient-specific models to study the influence of the plaque composition or morphology on the stenting outcome~\cite{welch_MechanicalInteractionExpanding_2016, kiousis_NumericalModelStudy_2007, corti_PlaqueHeterogeneityInfluences_2025,samant_ComputationalExperimentalMechanical_2021,chiastra_ComputationalReplicationPatientspecific_2016}. 
However, since the plaque compositions were not spatially resolved, they only assessed the influence of the individual plaque composition but not the pathoanatomical context.

In normal loading conditions, the spatially resolved plaque composition was included in the models to assess the plaque vulnerability~\cite{mastrofini_ImpactResidualStrains_2024,curcio_3DPatientspecificModeling_2023,warren_AutomatedFiniteElement_2022,akyildiz_EffectsIntimaStiffness_2011,corti_PredictingHighRiskPatients_2024}.
In~\cite{kadry_PlatformHighfidelityPatientspecific_2021}, the influence of the specific plaque morphology on the mechanical response was studied.  
However, these studies did not assess the mechanics during or after stenting. 
For the simulation of PCI, the spatially resolved plaque morphology was considered in~\cite{gharaibeh_CoregistrationPrePoststent_2020,zhao_PatientspecificComputationalSimulation_2021,poletti_DigitalTwinCoronary_2022,antonini_ComputationalWorkflowModeling_2024}, focusing on the realistic stent implantation through pre- and post-stenting imaging validation. 
In~\cite{dong_OpticalCoherenceTomographyBased_2020}, the impact of heavy calcifications on the stenting outcome was studied. 
The effects of separate plaque components in carotid arteries were analyzed in~\cite{fan_PlaqueComponentsAffect_2016}.
However, the influence of specific morphological plaque features on the stress state during and after PCI and the clinical outcome regarding ISR has not been assessed yet and is studied in this paper.

In previous computational modeling studies, the plaque composition and morphology information was often retrieved from histology~\cite{nieuwstadt_InfluenceAxialImage_2013,corti_PlaqueHeterogeneityInfluences_2025,holzapfel_ChangesMechanicalEnvironment_2005}, magnetic resonance imaging (MRI)~\cite{kiousis_NumericalModelStudy_2007,fan_PlaqueComponentsAffect_2016}, or OCT~\cite{samant_ComputationalExperimentalMechanical_2021,chiastra_ComputationalReplicationPatientspecific_2016,dong_OpticalCoherenceTomographyBased_2020,antonini_ComputationalWorkflowModeling_2024,zhao_PatientspecificComputationalSimulation_2021,poletti_DigitalTwinCoronary_2022,gharaibeh_CoregistrationPrePoststent_2020}. 
However, histology and MRI cannot assess coronary arteries in-vivo. 
OCT requires mapping to a centerline using angiography data, or it can only be used in artificially straightened artery models. 
Coronary computed tomography angiography (CCTA) is an emerging non-invasive imaging technology that enables the fast acquisition of a high-resolution, 3D image of the coronary arteries. 
Most notably, it can quantify of the plaque extension, volume, and composition~\cite{andreini_PreproceduralPlanningCoronary_2022}.
The tissue absorbance determines the grey value of the CCTA image, so-called \revb{Hounsfield units} (HU), which can be used to distinguish the components of coronary atherosclerotic lesions~\cite{shaw_SocietyCardiovascularComputed_2021,nieman_StandardsQuantitativeAssessments_2024}. 
The mapping from the HU value to the distinct component can be performed using fixed thresholds or adaptive algorithms, such as Gaussian \revb{m}ixture \revb{m}odel (GMM) clustering~\cite{gupta_GaussianmixturebasedImageSegmentation_1998}.
The latter method accounts for the CCTA limitation of different kilovoltage settings and acquisition parameters, which makes a unique comparison of plaque characteristics with fixed thresholds problematic~\cite{shaw_SocietyCardiovascularComputed_2021}.
Threshold- or GMM-based clustering of CCTA-based plaque information correlates well with established plaque analysis techniques\revb{, as population-level validations suggest}~\cite{athanasiou_ThreedimensionalReconstructionCoronary_2016,klass_CoronaryPlaqueImaging_2010,voros_ProspectiveValidationStandardized_2011,degraaf_AutomaticQuantificationCharacterization_2013}.

This work analyzes \revb{how} specific plaque morphological features \revb{influence local mechanical states after stenting, and how these mechanical states may be related to ISR}. 
We use computational modeling of the stenting procedure to assess stresses on the vessel wall during and after PCI.
The individual plaque morphology is included in the computational models using a GMM-based mapping from CCTA imaging data. 
We present four \revb{representative} example lesions from clinical cases with different morphological features; the respective angiography and catheter laboratory data are used to set up the stenting simulations. 
\revb{We compare the stresses after PCI with the ISR locations observed in follow-up examinations for these example cases, exploring stress levels that may be relevant for ISR initiation.} 
\revb{We further examine how specific morphological plaque features influence the local stress distribution and identify features and locations associated with higher stresses in these illustrative examples.}


\section{Methods for patient-specific coronary artery modeling}
\label{sec:methods}

The computational models and methods to simulate PCI in patient-specific coronary arteries were proposed in the authors' previous publication~\cite{datz_PatientspecificCoronaryAngioplasty_2025}. 
In the following section, we introduce the inclusion of plaque composition information from CCTA imaging into the patient-specific models and summarize the simulation setup. 

\subsection{Plaque composition mapping from CCTA}
\label{sec:HUmapping}

\begin{figure}[h]
\centering
\includegraphics[width=\textwidth]{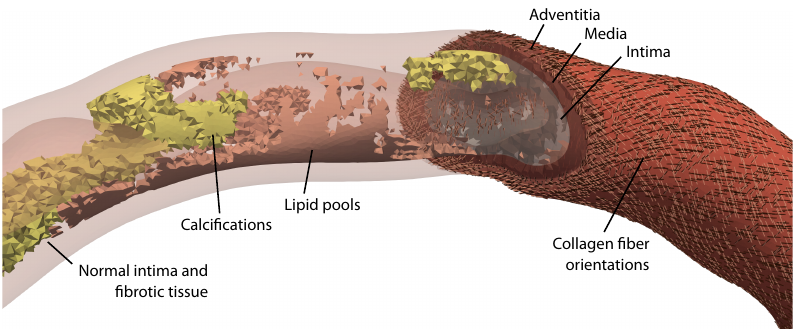}
\caption{Extract of a patient-specific coronary artery model. The fibrotic tissue and normal intima are not shown for a better overview.}
\label{fig:artery_all}
\end{figure}

The artery comprises three layers, i.e., the intima, tunica media, and adventitia, which are modeled as separate continua. 
Details on the model generation and materials are given in~\ref{app:modelgeneration}. 
The intima is further heterogeneously classified into the plaque components: lipid-rich tissue, fibrotic tissue, normal (healthy) intimal tissue, and calcifications. 
Figure~\ref{fig:artery_all} shows the components of the patient-specific artery model for an exemplary section. 

\begin{figure}[h]
\centering
\includegraphics[width=\textwidth]{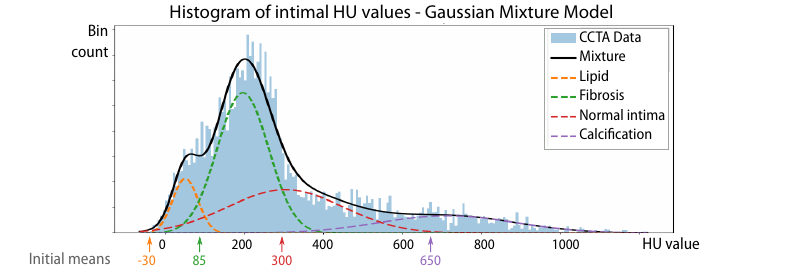}
\caption{\revb{GMM} for the plaque tissue of an exemplary lesion. Light blue: CCTA data; Gaussians of the different components and their initial means, i.e., the pre-defined mean values for each component, are shown in yellow, green, red, or purple for lipid-rich tissue, fibrosis, normal intimal tissue, or calcifications, respectively. Black line: Mixture of all components. }
\label{fig:GMM_ex}
\end{figure}

We assume that the HU value of a plaque component is not necessarily within one fixed range but can instead be represented by a Gaussian distribution. 
The lesion is composed of a linear combination of these distributions, i.e., the Gaussian distributions of the components can overlap. 
This overlapping can be attributed to the fact that the plaque components transform continuously on a microstructural level. 
To cluster the components with adaptive thresholds, we use a \revb{GMM} segmentation.
A GMM is a probabilistic model that assumes the data is a mixture of Gaussian distributions with unknown parameters~\cite{gupta_GaussianmixturebasedImageSegmentation_1998}.
The GMM is constructed from the HU data of the intima and initialized with the typical values for the plaque components taken from~\cite{nieman_StandardsQuantitativeAssessments_2024}. 
Each HU value is assigned to a component using the GMM mapping. 
Figure~\ref{fig:GMM_ex} shows the GMM on the histogram plot of the intimal tissue of an exemplary lesion.

\begin{figure}[h]
\centering
\includegraphics[width=\textwidth]{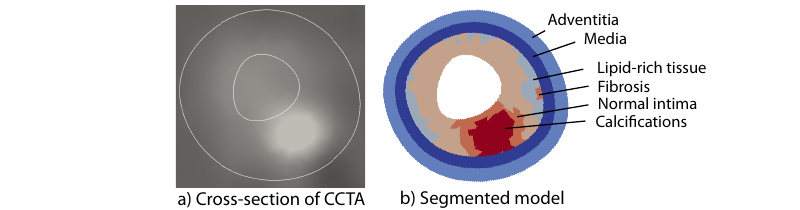}
\caption{Plaque segmentation of an example lesion. a) Cross-section of the CCTA image orthogonal to the vessel centerline with initially segmented outlines. b) Same slice in the clustered model with assigned plaque components}
\label{fig:crosssec_mat}
\end{figure}

This approach considers the whole 3D geometry for the plaque segmentation.
Instead of viewing the cross-section slices individually, the segmentation also contains information about the axial dispersion of the plaque. 
The 3D modeling of the plaque enables a realistic representation of the plaque morphology, which is crucial for assessing the influence of specific plaque features on local stresses.
\revb{The resulting plaque classification represents an approximate, morphology-level segmentation, but does not resolve plaque microstructure at the voxel scale.}
The segmentation of an exemplary lesion is shown in Fig.~\ref{fig:crosssec_mat} on a cross-sectional slice. 
Details on the tools and specifications of the GMM-based mapping are given in~\ref{app:GMM}.

\subsection{Percutaneous coronary intervention simulation}

\reva{The simulations follow a standardized workflow: patient-specific coronary artery geometries are segmented from CCTA, the stent is positioned according to angiographic guidance, and balloon inflation is simulated to deploy the stent within the artery.}
The structur\revs{al} mechanics simulations employ finite element methods based on the underlying physics equations, i.e., balance equations, constitutive relations, and the respective boundary conditions, and are done in our in-house open-source multiphysics high-performance code 4C~\cite{4c_4CComprehensiveMultiPhysics_2024}. 
\reva{As the focus is on stress states after stenting rather than transient dynamics, we} use a quasi-static pseudo-time discretization and apply the loads in small \reva{\mbox{(pseudo-)}}time steps.
For details on the simulation methods and background, the reader is referred to \reva{\ref{app:modelgeneration}} and~\cite{datz_PatientspecificCoronaryAngioplasty_2025}. 
Here, we only summarize the main points. 

\begin{figure}[h]
    \centering
    \includegraphics[width=\textwidth]{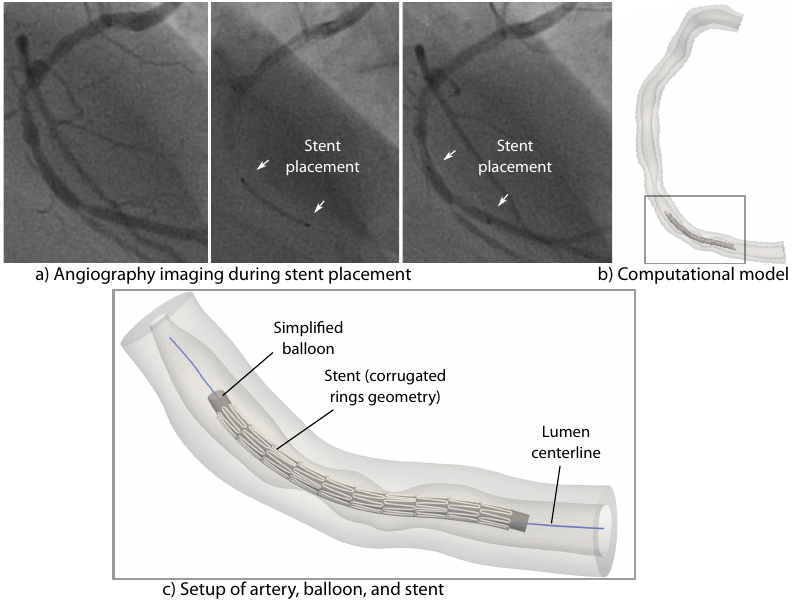}
    \caption{Stent placement in the computational model guided by the angiography imaging during the intervention}
    \label{fig:stentposangio}
\end{figure}

\reva{The stent is modeled using reduced-dimensional Simo-Reissner beam elements, which are well suited to represent the slender struts and bending-dominated deformation mechanisms governing the radial expansion.}
\reva{Elasto-plastic behavior is included to capture the permanent deformation responsible for the vessel support after stenting.}
The stent geometry is based on the ``corrugated rings'' design (see, e.g.,~\cite{_BiomatrixAlphaBiosensors_}), and modeled with~\cite{beammeauthors_BeamMeGeneralPurpose_}.
It is shown in Fig.~\ref{fig:stentposangio}c).
The edges are rounded, resulting in a more realistic expansion shape compared to~\cite{datz_PatientspecificCoronaryAngioplasty_2025}.
\reva{The balloon catheter modeled using a three-dimensional finite element formulation and is represented by a simplified, thin-walled elastic model that reproduces realistic inflation behavior while substantially reducing computational cost.}
The stenting device\reva{, i.e., balloon and stent,} is positioned based on the coronary angiography imaging data, shown exemplarily for one lesion in Fig.~\ref{fig:stentposangio}.
The initial configuration of the stenting device is assumed to be stress-free. 

Furthermore, recently developed methods for the mixed-dimensional\reva{, frictionless} contact interaction between the stenting device and the artery are employed~\cite{steinbrecher_ConsistentMixeddimensionalCoupling_2025}. 
We prescribe a pressure boundary condition to the inner surface of the balloon, leading to its radial expansion. 
The ends of the balloon and the cut-off surfaces of the artery model are fixed using spring-dashpot boundary conditions. 
The stent ends are fixed to allow only radial expansion, preventing the stent from moving in the axial direction.


\section{Patient-specific simulations}
\label{sec:numericalexamples}

In the following, we present the computational simulation of the stenting procedure on four lesions based on clinical cases from the TUM University Hospital German Heart Center. 
We use the coronary angiography and catheter laboratory data to assess the patient demographics, lesion specifications, and details about the type of intervention and outcome. 
The CCTA data is the basis for the model geometry and morphology, as detailed in Section~\ref{sec:methods}.
We assess the stress distributions in the artery wall during the maximum balloon expansion and the remaining stresses after the stenting.

\subsection{Patient selection and medical data}

\begin{figure}[h]
\centering
\includegraphics[width=\textwidth]{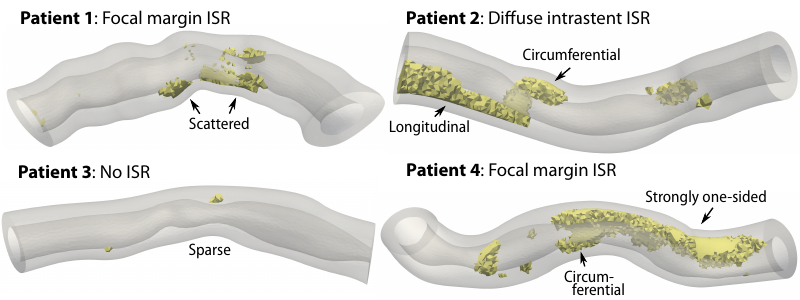}
\caption{Lesion sections of sample patients with calcification patterns (yellow).}
\label{fig:patientsHU}
\end{figure}

We selected the example lesions based on their specific morphological plaque features\revb{, to illustrate the influence of a different range of features relevant to the hypothesis.}
For all assessed lesions, coronary artery disease was diagnosed and treated with coronary angioplasty and stent implantation. 
Follow-up angiographic imaging revealed whether ISR had occurred in the stented region and if it was focal (between stents, at the stent edges, or the stent body), multifocal, or diffuse. 
Table~\ref{tab:patients} in~\ref{app:patientcharacteristics} summarizes the patient demographics, diagnosis, prior cardiovascular risk factors, treatment in the catheter laboratory, and the outcome at the follow-up after 6-12 months. 
For this study, we chose lesions that share a diagnosis (CAD in at least one right coronary artery segment) and treatment (angioplasty with stent implantation) for better comparability.
Figure~\ref{fig:patientsHU} shows the four selected lesions, highlighting the specific calcification characteristics and the clinical outcome at follow-up. 

\subsection{Stress distribution after the intervention} 
\label{sec:resultsstressdist}

PCI with stent implantation is simulated as described in Section~\ref{sec:methods}.
We assess the mechanical state after the PCI, i.e., after the balloon withdrawal. 
Concisely, we review the first principal stresses, i.e., the stress values in the direction of the maximum \revb{tensile} stress, as relevant mechanical value for vascular growth and remodeling~\cite{humphrey_VascularAdaptationMechanical_2008}. 
Although tissue strains are also relevant for adaptation, we focus on stresses as they inherently account for both mechanical loading and material properties. 
\ref{app:angios} shows the stent placement, final stented artery, and a comparison of high-stress locations with the clinical follow-up angiography results in detail for each patient. 

\begin{figure}[h]
    \centering
    \includegraphics[width=\textwidth]{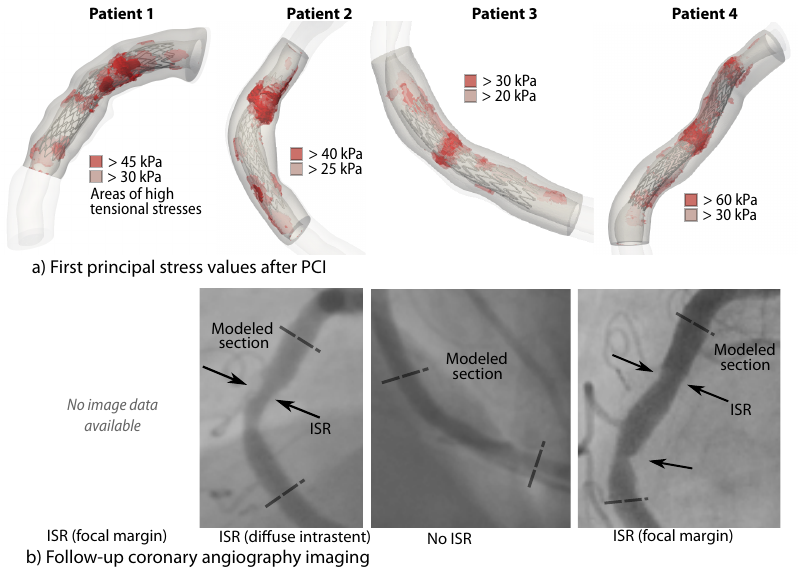}
    \caption{Comparison of a) regions of high \revb{tensile} stress after PCI simulation and b) angiography images at follow-up}
    \label{fig:summaryISR}
\end{figure}

At the maximum balloon expansion, we observe peak \revb{tensile} stresses of \unit[120]{kPa}.
After the balloon withdrawal, the \revb{tensile} stresses peak at around \unit[30-60]{kPa}.
Figure~\ref{fig:summaryISR} summarizes the high-stress and ISR locations in the angiography for the four example lesions. 
The threshold values for the simulation results are determined based on the overall stress values: The first threshold (light red) shows the top 20\% of the remaining stresses after PCI in the respective numerical example; the second threshold shows the top 5\% (dark red).
For Patient~1, no follow-up imaging was available, but only the clinical classification as focal margin ISR. 
From the remaining images, we observe \revb{that} high-stress locations after the stent simulation \revb{tend to occur at} ISR regions \revb{as} seen in the angiography images. 
The simulation of Patient~3 shows no significant stress elevations in the stented area; the respective follow-up angiography also shows no signs of ISR. 
The simulated results comply well with the lumen gain, as seen in the angiography. 

The areas of peak stress highly depend on the plaque morphology. 
In Patient~1, the lesion consists of mainly fibrotic and normal intima tissue with a small amount of scattered calcifications.
The highest stresses are reached at the stent boundaries and in the most stenotic section. 
Patient~2 shows exceptionally high stresses in the proximal area, where a specific calcification pattern switches from a longitudinal shape to winding around the lumen. 
Additionally, elevated stress levels can be observed at the distal stent boundary. 
In the mainly fibrotic lesion of Patient~3, which did not show signs of ISR in the clinical follow-up, we simultaneously observed only mildly elevated stresses around the narrowest lumen area. 
Patient~4 shows high stresses at the stent boundaries and in the middle section, where the calcification pattern forms a circumferential pattern around the lumen.

\subsection{\revb{ISR and local mechanical} stresses}
\label{sec:regionalanalysis} 

\begin{figure}
    \centering
    \includegraphics[width=\textwidth]{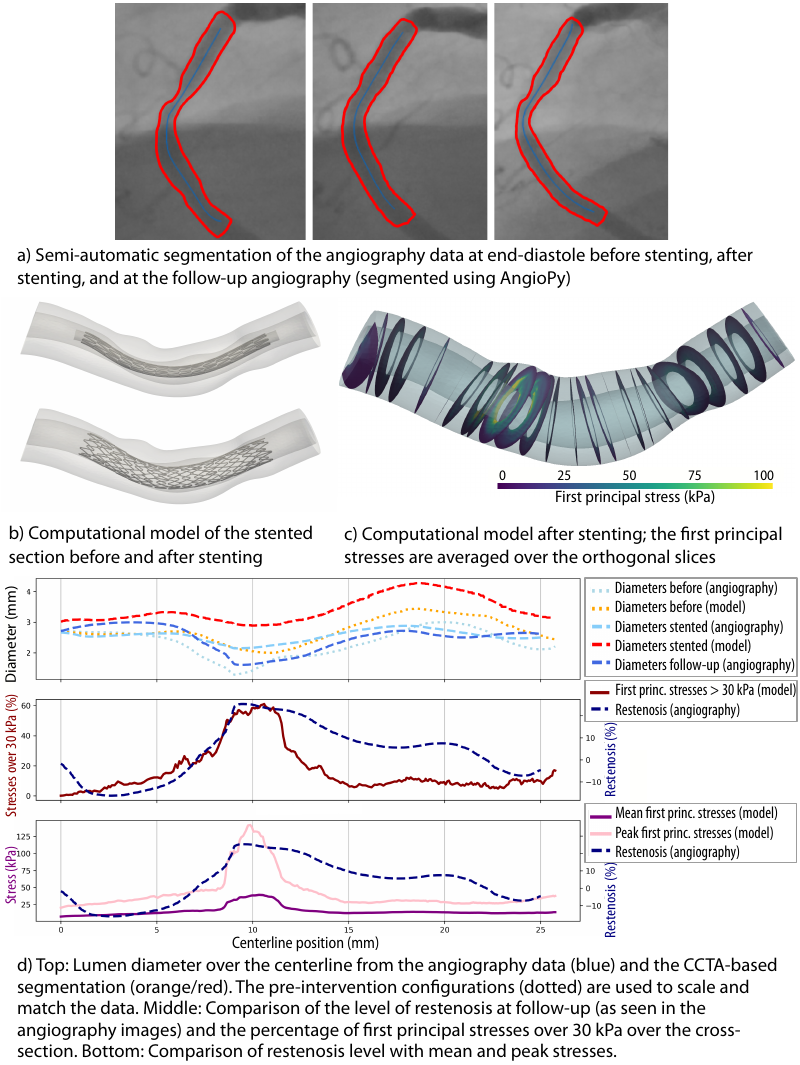}
    \caption{\reva{Regional analysis of vessel diameters and stresses after stenting for Patient~2}}
    \label{fig:regionalanalyis}
\end{figure}

We present a regional analysis of the artery diameters before stenting, after stenting, and at the follow-up angiography to \revb{assess how} the \revb{tensile} stresses along the centerline \revb{relate to} the amount of restenosis \revb{in the example cases}. 
For this analysis, we segment the end-diastolic frames of the angiography data using the open-source deep learning tool AngioPy, which was trained for coronary segmentation~\cite{mahendiran_AngioPySegmentationOpensource_2025}, see Fig.~\ref{fig:regionalanalyis}a).
Since no reference scale is available in the angiography images, we scale the segmented diameters to the diameter proximal to the stented region, assuming that the diameter stays approximately unaltered in this area. 
We approximate the percentage of regional restenosis using the diameter data after the stenting and at the follow-up angiography (dashed \reva{blue} lines). 
To match the datasets to the same section of the centerline, the maximum inscribed sphere diameter of the CCTA-based model and the diameters of the angiography data prior to the intervention are used, see Fig.~\ref{fig:regionalanalyis}d) (dotted lines).
\reva{The dashed red line shows the diameters of the model after stenting, also approximated using the maximum inscribed sphere diameter.}
However, a quantitative comparison of CCTA \reva{and hence, a CCTA-based model, with} angiography data is difficult due to the different image acquisition techniques. 
\reva{The larger post-stenting diameters predicted by the model may partly result from the lack of absolute scale in the angiographic images, and may additionally reflect a slight overestimation of lumen gain by the simulation.}

For the simulated model, we integrate the first principal stresses of each plane perpendicular to the centerline and divide them by the surface area to obtain the mean stress as a scalar value for each centerline point. 
The integrated stresses along the centerline are \revs{scaled in the same way as the diameters.}
We use PyVista for this analysis~\cite{sullivan_PyVista3DPlotting_2019}. 
Exemplary slices are shown in Fig.~\ref{fig:regionalanalyis}c). 
The mean local mechanical stresses in the stented region \revb{tend to be higher in regions with a high amount of ISR.}
We calculate a Pearson correlation coefficient of 0.54 for the data of the three example patients combined. 
\revb{Per-patient correlations can be found in \ref{app:perpatientcorr}.}
\revb{To quantify the statistical uncertainty, we estimate the 95\% confidence interval (CI) and empirical p-values of the slice-wise Pearson correlation.}
\revb{Because measurements along the vessel centerline exhibit strong spatial autocorrelation, we use a block bootstrap method to obtain the confidence intervals.} 
\revb{The conservative block size of 100 samples is chosen based on the empirical autocorrelation to avoid underestimating uncertainty.}
\revb{The sample size is then reduced from 910 slices to an approximate effective sample size of 9.}
\revb{For the mean stresses, the observed correlation is r = 0.54 (95\% CI: 0.39-0.71, p = 0.58).}
\revb{However, the high empirical p-value indicates substantial uncertainty, and the present results are insufficient to establish statistical significance.}
The peak stresses of each plane perpendicular to the centerline (shown in Fig.~\ref{fig:regionalanalyis}d), bottom) \revb{shows a correlation of r = 0.55 (95\% CI: 0.41-0.74, p = 0.62) with the amount of ISR}.
The $95^{\text{th}}$ percentile of the \revb{tensile} stresses is used instead of the absolute maximum to reduce the impact of local singularities and numerical outliers. 

\begin{figure}
    \centering
    \includegraphics[width=\textwidth]{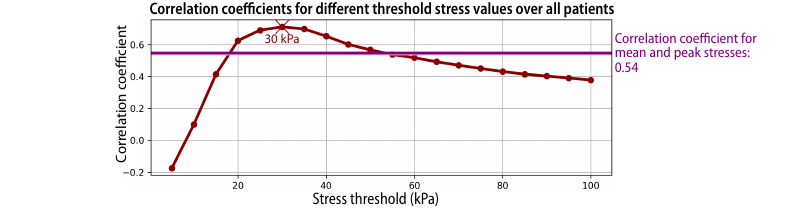}
    \caption{Correlation coefficients for different stress thresholds and for the mean and peak stresses (purple line) over all patients.}
    \label{fig:coeff_corr_all}
\end{figure}

Furthermore, we analyze the correlation between the tissue area experiencing \revb{tensile} stresses above a certain stress threshold and the amount of restenosis \revb{for the three sample lesions with follow-up data}. 
For each slice along the centerline, we compute the ratio of tissue with stresses above a certain stress threshold to the total slice area. 
We perform this analysis with different stress thresholds between 5 and \unit[100]{kPa} using the data of the three example patients combined.
\revb{This threshold sweep is intended as an exploratory, hypothesis-generating analysis rather than an inferential procedure.}
Figure~\ref{fig:coeff_corr_all} shows the \revb{Pearson} correlation coefficient for different stress threshold values.
The correlation of stresses above a certain threshold with the restenosis amount is highest for \unit[30]{kPa} \revb{as determined in a post-hoc analysis.}
\revb{For the \unit[30]{kPa} threshold, the observed correlation is r = 0.71 (95\% CI: 0.52-0.85, p = 0.51).}
\revb{This indicates that regions exceeding this stress threshold tend to coincide with local restenosis.}
\revb{However, the empirical p-values in this analysis indicate high uncertainty of the resulting correlation coefficients.}

\subsection{Calcification patterns and stress distributions}
\label{sec:calcpatterns}

We investigate calcification patterns in the cross-sectional view and their influence on the stress distribution post-PCI. 
In \ref{app:patspecresults}, the \revb{tensile} stress distributions are presented at several cross-sectional views during and after the PCI simulation for all four example lesions. 
In this section, we highlight two calcification patterns in the example lesions and assess their influence on the specific stress distribution in the cross-section.

\begin{figure}[h]
    \centering
    \includegraphics[width=\textwidth]{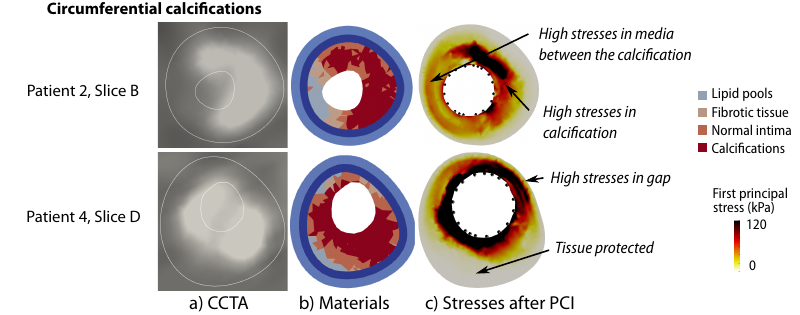}
    \caption{Circumferential calcification pattern a) as seen in the coronary CT angiography image, b) the material distribution in our model and c) the resulting first principal stress distributions after PCI}
    \label{fig:summaryresults_circ}
\end{figure}

We define a circumferential calcification pattern as a calcification of more than 180° in the cross-sectional view. 
In the example lesions, circumferential calcification patterns occur in Patients 2 and 4. 
Figure~\ref{fig:summaryresults_circ} shows two extracted cross-sections from Patients 2 and 4 showing the circumferential calcification pattern. 
We observe peak values of over \unit[120]{kPa} (black areas) near the circumferential calcifications. 
The locations of the cross-sections in the respective lesions are given in~\ref{app:patspecresults}. 
In the neighboring non-calcified tissue, which is in the two presented cross-sections composed of lipid-rich tissue or fibrotic tissue, the stresses in the intima and media are significantly elevated. 
The stresses in the calcified parts are also high.
However, calcifications possess high mechanical strength and are not susceptible to ISR, as they lack proliferative cells. 
The media and adventitia outside calcifications are protected against high stresses and strains.

\begin{figure}[h]
    \centering
    \includegraphics[width=\textwidth]{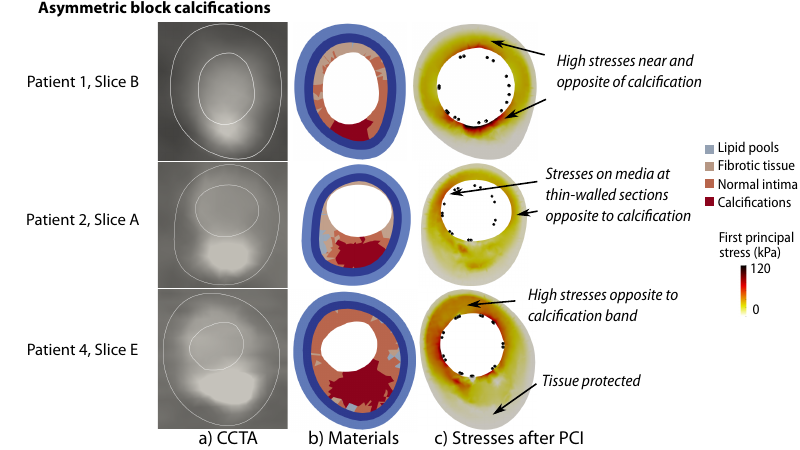}
    \caption{Asymmetric block calcifications a) as seen in the coronary CT angiography image, b) the material distribution in our model and c) the resulting first principal stress distributions after PCI}
    \label{fig:summaryresults_asym}
\end{figure}

The second specific calcification pattern is the asymmetric block calcification.
This pattern in the cross-section can also stem from a longer calcification band along the length of the artery.
This pattern occurs in the example patients 1, 2, and 4. 
\ref{app:patspecresults} shows the location of the specific cross-sections and the respective 3D model view. 
Similarly to the circumferential pattern, Fig.~\ref{fig:summaryresults_asym} shows the cross-sections of the respective model and the stress distributions after the PCI simulation. 
Near asymmetric block calcifications, we observe peak values of approximately \unit[60]{kPa}.
Our results suggest higher stresses near and opposing the block calcification \revb{in the analyzed cases}.
Especially in thin-walled sections on the opposite side, the intima and media experience higher stresses. 
The tissue behind the calcification seems protected from higher stresses by the calcification. 


\section{Discussion}

This study investigated the influence of specific plaque morphological features on the local mechanical states during and after PCI with stent implantation and \revb{how these mechanical states relate to the occurrence of ISR in the examined cases}. 
Using computational modeling and simulation on four relevant example lesions, \revb{we observed that regions of elevated stress coincided with the areas where ISR was detected in the follow-up angiography}. 
Furthermore, we have investigated how different plaque morphological features, such as circumferential or asymmetric block calcifications, influence the stress distribution after the stent implantation. 
The results of this study suggest that the mechanical state during and after PCI, such as stresses on the vessel wall, \revb{may provide a link between} specific plaque morphological features \revb{and the amount of ISR observed in these cases}.
\reva{Notably, the plaque morphologies that were observed in this study to influence the stress distribution, i.e., circumferential or block calcifications, are also known to promote incomplete stent expansion or malapposition, which themselves contribute to ISR.}
\reva{In this context, the elevated and heterogeneous stress distributions observed in the present simulations should be interpreted as part of a biomechanical mechanism linking lesion morphology, stent-vessel interaction, and ISR, rather than an isolated causal mechanism.}
\reva{Because local stresses reflect the combined effects of plaque morphology, stent expansion, vessel mechanics, and stent-tissue interaction, they may serve as an aggregate biomechanical indicator of lesion characteristics associated with ISR.}
The present study \revb{provides mechanistic, hypothesis-generating insights and} lays the groundwork for consecutive studies with larger patient cohorts and statistical significance to \revb{more reliably quantify these relationships}.

\subsection{Stress values relevant for ISR}

The regional analysis in Section~\ref{sec:regionalanalysis} showed that average values of \unit[30-40]{kPa} were reached in the areas with a 20\% or more restenosis rate as determined from the angiography images. 
The computed correlation between the mean \revb{tensile} stress and the corresponding diameter reduction after PCI was 0.54 for all three patients combined.
A correlation coefficient of 0.55, was computed for the peak \revb{tensile} stresses with the diameter reduction after PCI.

A qualitative analysis was performed for the 3D stress state in Section~\ref{sec:resultsstressdist}. 
The results suggest that ISR \revb{was most prevalent} in areas with \revb{tensile} stress values of approximately \unit[40-60]{kPa}.
This higher threshold shows that the mean or peak stress values over cross-sectional slices are insufficient to determine the regions at risk for ISR, but the local values are important. 
Therefore, we assessed the influence of regions with stresses above a certain threshold on the initiation of ISR in a further analysis in Section~\ref{sec:regionalanalysis}. 
We systematically reviewed the correlation of areas with stresses above different thresholds with the ISR rate, assuming that excessive stresses outside of the normal physiological range trigger growth and remodeling processes that can lead to ISR. 
The analysis led to a correlation of 0.71, \revb{suggesting a} correlation for areas with stresses above \unit[20-40]{kPa}. 
\revb{These results indicate that there exists a stress threshold that is mechanically relevant for the onset of ISR}
rather than a direct correlation with the stress values.
Hence, the knowledge of locations at risk for stress peaks is important for ISR prediction.  

\revb{However, as demonstrated in Section \ref{sec:regionalanalysis}, the uncertainty of the reported correlation coefficients is high.}
\revb{Given the small effective sample size, spatial dependence, and post-hoc nature of threshold selection, this observation remains hypothesis-generating.}
\revb{Additional lesions and out-of-sample validation will be required to determine whether a consistent and universal critical stress threshold exists.}
\revb{A statistically robust assessment accounting for spatial dependence and inter-patient variability would further require a mixed-effects framework, which is not feasible in this study given the small sample size.}
\revb{Additionally, the critical stress }threshold is likely individual for different patients and lesions, as multiple factors are involved \revb{in the onset of ISR, such as biological, clinical, or device-related factors}. 
In the literature, stress thresholds have been proposed for the plaque rupture. 
For instance,~\cite{holzapfel_ComputationalApproachesAnalyzing_2014,kwak_BiomechanicalFactorsAtherosclerosis_2014} suggested values of \unit[300-500]{kPa}, aortic aneurysms rupture at around \unit[450-1000]{kPa}~\cite{humphrey_MechanicsMechanobiologyModeling_2012}.
However, rupture is a purely mechanical outcome governed by the mechanical state alone, i.e., when stress exceeds strength~\cite{humphreyVascularMechanobiologyHomeostasis2021}. 
In contrast, ISR involves mechanobiological processes, which are determined by the perceived state of cells and set homeostatic targets~\cite{humphreyVascularMechanobiologyHomeostasis2021}.
Hence, the initiation of ISR is likely governed by a broader range of factors, making it more difficult to pinpoint clear trigger thresholds.
\reva{The high uncertainty of the correlations additionally suggests that tensile stress alone is not sufficient to explain the occurrence of ISR.}
\reva{While regions of high tensile stress coincided with areas of ISR at follow-up, this association was not present for all ISR regions.}
\reva{This indicates that additional procedural, biological, or hemodynamic factors likely contribute to ISR in areas that are not associated with high tensile stresses in the shown examples.}
\reva{These observations support interpreting mechanical stress as a complementary contributor rather than a specific or standalone predictor of ISR. }

\subsection{Influence of plaque morphology on ISR}

The stress distributions in the cross-sectional slices highlight the impact of plaque morphology on local stress levels, as shown in Section~\ref{sec:calcpatterns}.
The stresses in the healthy, surrounding tissue of the cross-sections showing specific plaque characteristics, such as circumferential or asymmetric calcifications, were significantly elevated.
Peak values of over \unit[120]{kPa} near the circumferential calcification patterns and approximately \unit[60]{kPa} near asymmetric block calcifications are reached. 
These patterns have already been observed to be critical for ISR. 
Other areas, such as the media and adventitia outside of a block calcification, show significantly lower stresses.
The location of the peak stresses after PCI \revb{matched} the angiography data at follow-up, as shown in Section~\ref{sec:resultsstressdist}, and the documented type of ISR, i.e., focal margin ISR for Patients 1 and 4, diffuse intrastent for Patient 2, and no ISR for Patient 3.
The threshold values are chosen based on the top 5 or 20\% of the overall stress distribution of each numerical example, respectively.
\revb{These percentile-based thresholds are used to visualize high-stress regions within each case and to identify where peak stresses occur relative to the observed ISR locations. Cross-patient observations refer to absolute stress magnitudes.}
For Patient~3 (no ISR), \revb{absolute peak stresses} were lower than the other three examples, which showed signs of ISR in the follow-up angiography. 
We conclude that plaque morphological patterns are relevant to determining the ISR risk. 
Based on these findings, more extensive studies can be conducted to identify more plaque characteristics at risk for ISR.


\subsection{Strengths and limitations} 

This work assessed the mechanical state of the artery during and after coronary stent implantation, as it is an important factor of ISR that is so far mostly neglected in the literature. 
Other factors are involved in the development of ISR, such as denudation at the stent struts, hemodynamic alterations influencing the local intimal wall shear stresses, the type of antiproliferative drugs, and the progression and diffusion of inflammatory and growth factors through the artery wall. 
These factors have not been studied in this work, but they must be considered to predict ISR.
\revb{Accordingly, our findings should be interpreted as considering only one mechanical contributor rather than describing the full mechano-biological cascade leading to ISR.}
\revb{The present analysis can serve as a basis for future studies coupling structural mechanics with hemodynamics, drug transport, and biological response, such as~\cite{ranno_ComputationalModelCoronary_2025}.}

This study showed the correlation between plaque morphological features and ISR by assessing the mechanical state of four representative lesions \revb{(of which only three included follow-up data)}. 
\revb{Due to this small sample size, the results should be interpreted as hypothesis-generating. To sustain definitive statistical inference, more data is needed.}

\reva{Assessment of stent expansion was based on angiography, which cannot definitively exclude underexpansion or malapposition, particularly in calcified lesions.}
\reva{Intravascular imaging would be required to derive definite conclusions.}
\reva{While the comparison of CCTA-based model outputs with angiographic measurements is limited by modality-specific scaling and projection effects, the availability of longitudinal angiographic data represents a valuable opportunity for qualitative validation of patient-specific model predictions.}

\revb{The segmentation accuracy is an inherent source of uncertainty in image-based modeling.}
\revb{However, the main conclusions of our analysis rely on relative trends across stresses rather than absolute magnitudes, which are more robust to small geometric perturbations.}
\revb{Furthermore, the spatial resolution of the CCTA images and the typical segmentation variability of a few voxels (one voxel is approximately \unit[0.4]{mm}) are small compared to the characteristic geometric features, which are mostly larger than \unit[3-4]{mm}.}
\revb{Hence, we expect the segmentation uncertainty to have limited influence on the results.}

\revb{We focused on the first principal stress as a mechanistic and interpretable metric for vessel injury.}
\revb{Additional metrics, such as strain, strain energy density, stress gradients, and more sensitivity analyses (parameter sweeps, confidence bands) were beyond the scope of this exploratory study but represent important future directions.}

We chose the material parameters from experimental data reported in the literature. 
However, this study did not include an uncertainty quantification of these parameters, which would be an important step toward improving the accuracy of the stress estimates. 
\revb{Systematic sensitivity analyses with physiological variations of material properties, alternative boundary conditions, and the inclusion or exclusion of residual stresses would further strengthen the robustness and generalizability.}
\revb{Future studies should therefore assess the impact of such modeling choices by varying parameters and by reporting uncertainty bands for stress-based metrics and thresholds.}

\revb{As this study focuses on the relationship between selected plaque morphological features, mechanical states during stenting, and ISR locations, several aspects of the stent-artery modeling approach were not included in the present study, such as the exact stent geometry, stent crimping, or cardiac cyclicity.}
\revb{A detailed justification of the modeling assumptions, boundary conditions, dimension reduction, and material choices employed here has been provided previously in our methodological paper~\cite{datz_PatientspecificCoronaryAngioplasty_2025}, to which we refer the reader for a more detailed discussion.}
\revb{Cases were modeled using the clinically documented stent type and deployment parameters, to explore how distinct morphologies influence the spatial distribution of tensile stresses within each lesion under realistic PCI conditions.}
\revb{We did not control or stratify for procedural or device-related heterogeneity, e.g., stent type or balloon pressures, since this was not feasible given the small sample size.}
\revb{Therefore, the findings are restricted to lesion-specific spatial patterns and not to standardize or compare device- or procedure-specific effects across patients.}

\revb{In this work, we chose the mesh sizes based on mechanical considerations of the layered artery structure.}
\revb{The mesh was refined such that approximately one element through the thickness of each layer was ensured.}
\revb{Quadratic base functions were used to capture the stress gradients within the wall.}
\revb{Future work should include a systematic mesh sensitivity analysis to further quantify numerical sensitivity.}

We applied a GMM-based clustering of the HU values to distinguish between plaque components. 
\revb{In the absence of phantom calibration, inter-/intra-observer reproducibility analysis, or co-registration with intravascular imaging, the segmentation was used solely to define patient-specific morphological features for mechanical modeling rather than to establish histological ground truth.}
\revb{Furthermore,} this approach is a simplification, as it reduces the complex and continuous microstructural transitions to discrete categories. 
Therefore, a soft clustering of the plaque materials could be more realistic.


\subsection{Clinical relevance and future research} 

\revb{This study illustrated that cases with a comparable baseline clinical risk by standard descriptors and similar standard imaging descriptors (see Table~\ref{tab:patients} in \ref{app:patientcharacteristics}) can exhibit markedly different local tensile stress distributions, and that higher stresses were observed in regions that subsequently developed ISR.}
\revb{This suggests that biomechanical stress may serve as complementary information beyond standard imaging descriptors, motivating future validation in larger, statistically powered cohorts.} 
To assess this \revb{relationship}, we performed numerical simulations of the stenting procedure and utilized comprehensive clinical data, i.e., CCTA, coronary angiography, and interventional data before, during, and after PCI.
\revb{In future work, systematic validation of the relationship between lesion morphology, biomechanical stress states, and ISR in statistically powered cohorts could enable stress-informed interpretation of routinely available imaging, such as CCTA.}
\revb{Once validated, such approaches may help identify lesions with potentially elevated mechanical risk using standard imaging, therefore informing risk stratification and the selection of patients who may benefit from more extensive invasive assessment, such as intravascular imaging.}

\revb{We reported relative reocclusion as percent change from the initial lumen diameter, absolute calibration from quantitative coronary angiography or OCT was not available for the cases in this study, which may introduce a scaling bias.}
\revb{Confirmation with dedicated additional retrospective data, such as quantitative coronary angiography or OCT, would further strengthen the clinical validation and should be considered in future work.}

The morphological plaque features could be resolved in more detail using a soft clustering approach. 
Using GMM clustering, the probability of a distinct HU value belonging to a different material could be used as the basis of a constrained mixture of materials of the respective components, see e.g.~\cite{humphrey_ConstrainedMixtureModels_2021,cyron_HomogenizedConstrainedMixture_2016}. 
An inverse analysis with corresponding experimental data and microstructural insights on how the materials transform into each other would be beneficial to fine-tune this mapping. 

The approach introduced in~\cite{datz_PatientspecificCoronaryAngioplasty_2025} is applicable for any coronary artery section, independently of its degree of stenosis, plaque composition, and morphology. 
To further quantify the critical stress thresholds for ISR, we suggest a study with a larger number of lesions.
Additionally, a larger patient study should be used to derive statistically relevant correlations between the plaque morphology and \revb{tensile} stress distribution and eventually predict ISR.

\newpage

\appendix

\section{Model generation from imaging data}
\label{app:modelgeneration}

We create the patient-specific stenotic artery models, as detailed in~\cite{datz_PatientspecificCoronaryAngioplasty_2025}. 
The segmentation from CCTA imaging data is done using~\cite{updegroveSimVascularOpenSource2017}.
\revb{After manually defining the centerline of the vessel, the inner and outer vessel walls are segmented slice-wise using the machine learning-guided segmentation tool of SimVascular~\cite{updegroveSimVascularOpenSource2017}.}
\revb{From the segmented slices, the final geometry is warped along the centerline.}
\revb{A mesh smoothing step reduces small artifacts in the final geometry.}
\revb{Finally, the lumen geometry is compared to the coronary angiography imaging data and refined, where necessary.}
\revb{The final segmented artery is cut to include only the stented region and approximately \unit[5]{mm} of the adjacent vessel on each side.}

We assume that the pathological thickening of the wall affects only the intima. 
Hence, the media and adventitia are modeled with fixed thicknesses.
\revs{The layer thicknesses of media and adventitia are calculated based on the ratio to the outer vessel diameter, based on the experimentally derived ratios in~\cite{holzapfelDeterminationLayerspecificMechanical2005}.}
The remaining intimal tissue is further classified into specific plaque components, as described in Section~\ref{sec:HUmapping}. 
The artery material is modeled as (almost) incompressible, anisotropic, viscoelastic, with a non-linear stress-strain curve. 
We use a Neo-Hookean base material with exponential fiber contributions to describe the constitutive behavior \revb{of the media and adventitia}~\cite{gasser_FiniteElementModeling_2007,holzapfelDeterminationLayerspecificMechanical2005}. 
\revb{Plaque components in the intima are modeled using a simplified linear elastic formulation based on the data reported in~\cite{holzapfelDeterminationLayerspecificMechanical2005}.}
\revb{This simplification is motivated by the following considerations.}
\revb{First, the volume fraction of each plaque constituent is small compared to the mechanically dominant media and adventitia layers.}
\revb{In preliminary simulations, replacing the full anisotropic hyperelastic formulation with a linear elastic approximation resulted in negligible changes in the predicted stress fields within the stented region.}
\revb{Second, applying highly non-linear hyperelasticity to small, heterogeneous regions leads to local stiffness contrasts that cause numerical instabilities and impaired solver convergence during balloon expansion.}
\revb{The simplified formulation substantially improves numerical robustness while preserving the overall mechanical response relevant for the quantities analyzed in this study.}
The material models and parameters are specified in~\ref{tab:materialparams}.

\begin{table}[h]
\centering
\caption{Material parameters of the artery model}
\label{tab:materialparams}
\ra{0.9}
\resizebox{0.8\textwidth}{!}{
\begin{tabular}{@{}rrrrrrr@{}}\toprule
& \multicolumn{2}{c}{Base material} & \phantom{a} & \multicolumn{3}{c}{Fibers}\\
& $E$ (MPa) & $\nu$ (-) && $k_1$ (kPa) & $k_2$ (-) & $\phi$ (\degree) \\ \midrule
Adventitia & 0.016 & 0.45 && 5.1 & 15.4 & 56.3 \\ 
Media & 0.16 & 0.45 && 0.64 & 3.54 & 5.76\\ 
Normal intima & 0.16 & 0.45&&  &  & \\ 
Lipid-rich plaque & 0.08 & 0.45 &&  &  & \\ 
Fibrotic plaque & 0.16 & 0.45 &&  &  & \\ 
Calcification & 1.6 & 0.45 &&  &  & \\ 
\bottomrule
\end{tabular}}
\end{table}

\revb{The meshing of the artery is done using LNMMeshio and Gmsh~\cite{gebauer_LNMmeshio_2023,geuzaine_Gmsh3DFinite_2009}.}
\revb{A tetrahedral finite element discretization with quadratic basis function is employed, total element and node counts, as well as the mean mesh sizes are given in Table~\ref{tab:elecounts}.}

\begin{table}[h]
\caption{\revb{Elements, nodes and mesh size}}
\label{tab:elecounts}
\centering
\ra{0.9}
\begin{tabular}{@{}rrrrrrr@{}}\toprule
 & Elements & Nodes & Mesh size\\ \midrule
Patient 1 & 106,769 & 164,912 & 0.3 mm \\ 
Patient 2 & 156,465 & 237,191 & 0.35 mm \\ 
Patient 3 & 81,133 & 125,562 & 0.3 mm \\ 
Patient 4 & 204,179 & 309,916 & 0.3 mm \\ 
\bottomrule
\end{tabular}
\end{table}

\section{Gaussian \revb{m}ixture \revb{m}odel mapping}
\label{app:GMM}

For the GMM-based mapping of the plaque composition described in Section~\ref{sec:HUmapping}, the following tools and specifications are employed. 
We use the Scikit Learn library~\cite{pedregosa_ScikitlearnMachineLearning_} to learn the GMM from the HU data of the intima domain and cluster the data accordingly. 
Each data point, i.e., discrete HU value, is assigned to the Gaussian it most likely belongs to~\cite{gupta_GaussianmixturebasedImageSegmentation_1998}. 
We use the `k-means' method with pre-defined mean values to initialize the GMM. 
The means are chosen based on the typical HU ranges for the different plaque components, see e.g.~\cite{nieman_StandardsQuantitativeAssessments_2024} and the number of plaque components is specified to four. 
Additionally, we use SimpleITK~\cite{lowekamp_DesignSimpleITK_2013} as an interface from the medical imaging data to the mapping. 

\newpage

\section{Patient characteristics}
\label{app:patientcharacteristics}

The patient and interventional data for the examples in Section~\ref{sec:numericalexamples} are summarized in Table~\ref{tab:patients}. 

\begin{table}[h]
\caption{Patient and interventional data. CV: cardiovascular, RCA: right coronary artery, LV: left ventricular, CVD: cardiovascular disease, ISR: in-stent restenosis}
\label{tab:patients}
\centering
\ra{0.9}
\resizebox{\textwidth}{!}{
\begin{tabular}{ll}\toprule
\textbf{Patient 1} & Female, 58y.\\ 
$\;$ \textit{Diagnosis} & Proximal RCA severely stenosed, stable angina pect \\ 
$\;$ \textit{CV risk factors}  & Smoker, arterial hypertension, hypercholesterolemia \\ 
$\;$ \textit{Treatment}  & (Successful) stenting of the RCA \\ 
$\;$ \textit{Follow-up}  & Restenosis (focal margin) \\ \midrule
\textbf{Patient 2} & Male, 71y.\\ 
$\;$ \textit{Diagnosis} &  Medial RCA severely stenosed, exertional dyspnea  \\ 
$\;$ \textit{CV risk factors}  &  Adiposity, arterial hypertension, hypercholesterolemia  \\ 
$\;$ \textit{Treatment}  &  Stenting of RCA, post-dilatation was necessary in proximal section  \\ 
$\;$ \textit{Follow-up}   & Restenosis (diffuse intrastent) \\ \midrule
\textbf{Patient 3} & Female, 68y.\\ 
$\;$ \textit{Diagnosis} & Triple-vessel coronary artery disease, reduced LV function\\ 
$\;$ \textit{CV risk factors}   & Prior interventions, hypertension, adiposity, hypercholesterolemia\\ 
$\;$ \textit{Treatment}  &  Four stents in RCA, post-dilatation in some areas \\ 
$\;$ \textit{Follow-up}  & Good result in stent area, no recurrent treatment necessary \\ \midrule
\textbf{Patient 4} & Male, 61y.\\ 
$\;$ \textit{Diagnosis} &  Severe stenosis of proximal, medial and distal RCA\\ 
$\;$ \textit{CV risk factors}   & Family history of CVD, hypercholesterolemia\\ 
$\;$ \textit{Treatment}  & Three stents in RCA, pre- and post-dilatation \\ 
$\;$ \textit{Follow-up}  & ISR in RCA (focal margin) \\
\bottomrule
\end{tabular}}
\end{table}

\section{Stent placement and correspondence to angiography data}
\label{app:angios}

\begin{figure}[h]
\centering
\includegraphics[width=\textwidth]{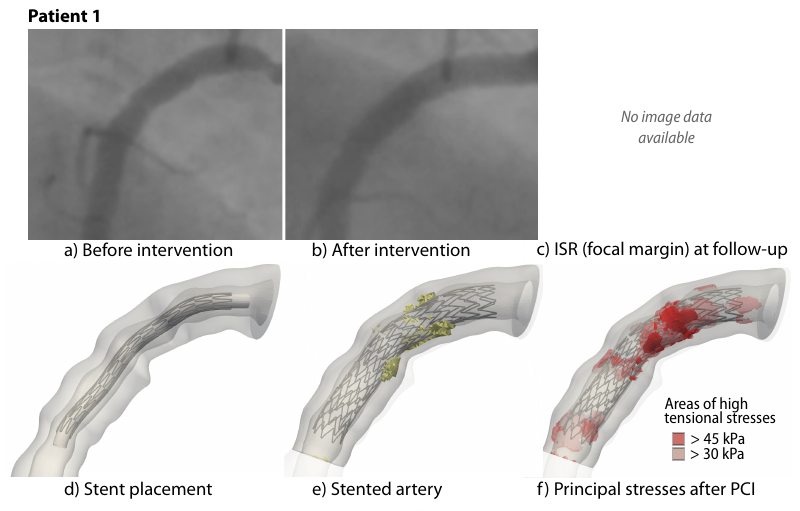}
\caption{\revs{Stent placement and corresponding angiography data for Patient 1}}
\label{fig:pat4_whole}
\end{figure}

\begin{figure}[h]
\centering
\includegraphics[width=\textwidth]{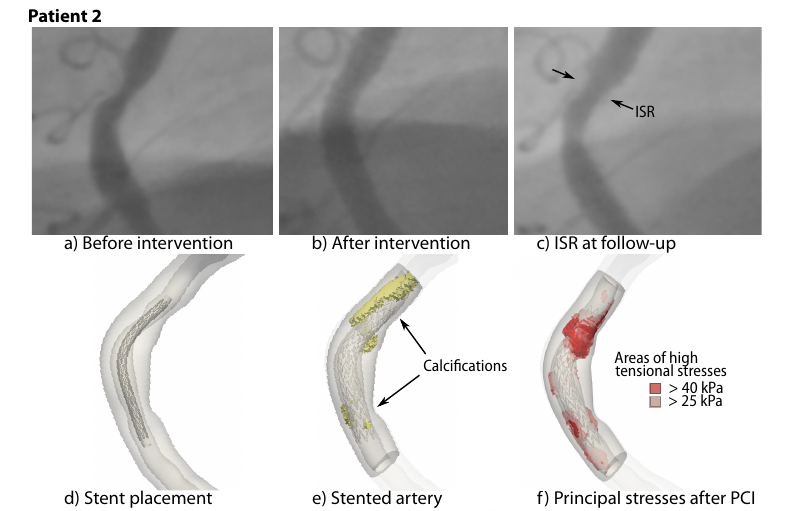}
\caption{\revs{Stent placement and corresponding angiography data for Patient 2}}
\label{fig:pat10_whole}
\end{figure}

\begin{figure}[h]
\centering
\includegraphics[width=\textwidth]{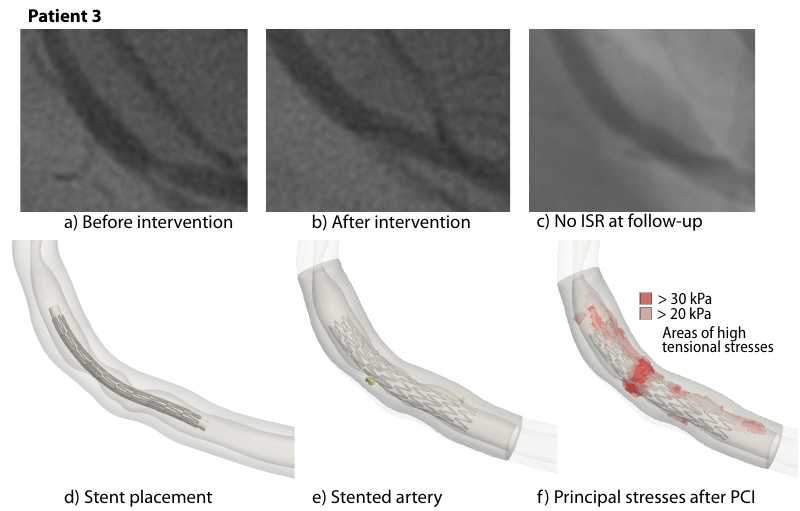}
\caption{\revs{Stent placement and corresponding angiography data for Patient 3}}
\label{fig:pat12_whole}
\end{figure}

\begin{figure}[h]
\centering
\includegraphics[width=\textwidth]{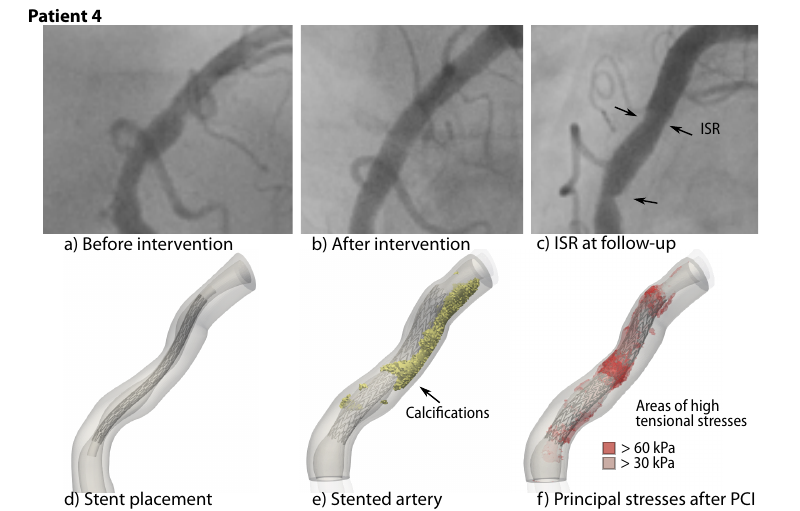}
\caption{\revs{Stent placement and corresponding angiography data for Patient 4}}
\label{fig:pat14_whole}
\end{figure}

In the following, we compare the simulation results \revs{of Patient 1, 2, 3, and 4} to the angiography imaging data\revs{, see Fig.~\ref{fig:pat4_whole}, \ref{fig:pat10_whole}, \ref{fig:pat12_whole}, and \ref{fig:pat14_whole}, respectively}. 
We select single frames during the diastole from the respective angiography data. 
For each patient, Subfigures a) and d) show the configuration before the intervention. 
Subfigures b) and e) show the angiography image after the intervention and the simulated stented artery. 
Subfigures c) and f) compare the follow-up angiography with the maximum stresses after the intervention. 


\clearpage

\section{\revb{Per-patient correlation coefficients}}
\label{app:perpatientcorr}

\revb{Table~\ref{tab:corrs} shows the per-patient correlation coefficients with 95\% CI and p-values.}

\begin{table}[h]
\caption{\revb{Per-patient correlation coefficients (r) with 95\% CI and p-values for mean, peak, and thresholded (\unit[30]{kPa}) stress metrics.}}
\label{tab:corrs}
\centering
\ra{0.9}
\begin{tabular}{@{}llccc@{}}
\toprule
 &  & Patient 2 & Patient 3 & Patient 4 \\ 
\midrule
\multirow{3}{*}{Mean stresses}
 & r  & 0.78 & -0.29 & 0.34 \\
 & 95\% CI & 0.73-0.91 & -0.5-0.86 & -0.35-0.67 \\
 & p  & 0.63 & 0.85 & 0.57 \\
\midrule
\multirow{3}{*}{Peak stresses}
 & r  & 0.72 & -0.31 & 0.29 \\
 & 95\% CI & -0.31-0.89 & -0.51-0.74 & -0.31-0.57 \\
 & p  & 0.74 & 0.89 & 0.51 \\
\midrule
\multirow{3}{*}{Thresholded}
 & r  & 0.75 & -0.37 & 0.57 \\
 & 95\% CI & 0.01-0.92 & -0.53-0.88 & -0.15-0.71 \\
 & p  & 0.81 & 0.55 & 0.59 \\
\bottomrule
\end{tabular}
\end{table}

\clearpage

\section{Patient-specific stress results during and after PCI}
\label{app:patspecresults}

\subsection{Stress distribution in cross-sections}

In the following, we present the complete results for the stress distributions in cross-sectional slices of the four numerical examples.

\begin{figure}[h]
\centering
\includegraphics[width=\textwidth]{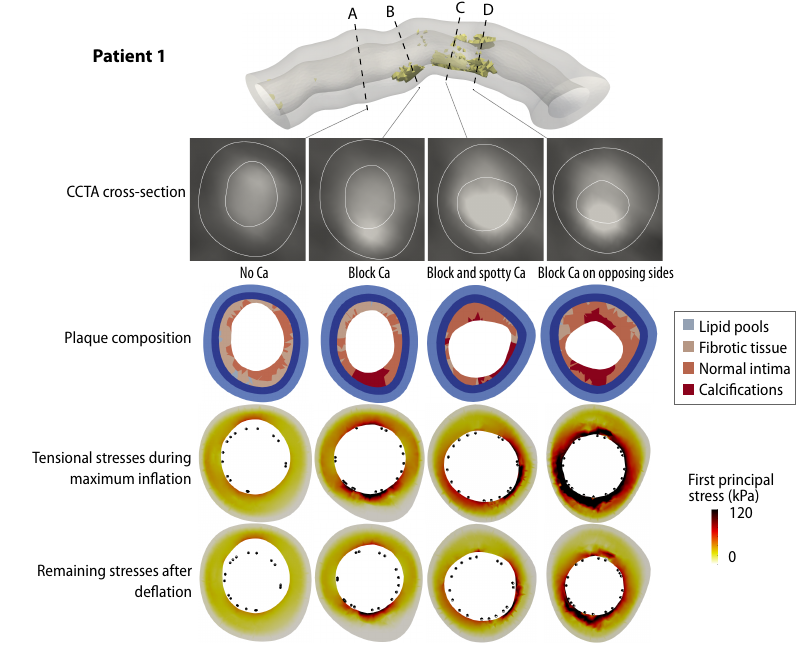}
\caption{First principal stress values in selected cross-sections during and after PCI; Patient 1. The plaque morphology is shown in the \revb{CCTA} cross-sections and in the model. The stent cross-section is shown as black dots. Ca: Calcification.} 
\label{fig:pat4_crosssec}
\end{figure}

Patient 1 shows the following plaque characteristics in the cross-sections: one-sided block calcification (Slice B), small block calcification with scattered calcified particles (Slice C), opposing calcifications (Slice D), see Fig.~\ref{fig:pat4_crosssec}. 
During maximum balloon inflation (Fig.~\ref{fig:pat4_crosssec}, third row), stress peaks can mainly be observed near the calcifications. 
High stresses occur in the media and adventitia, especially in the thin-walled area next to the larger calcifications in Slice D.
After PCI, the stress values are reduced; however, the spatial distribution is similar to the maximum pressure state. 

\clearpage

\begin{figure}[h]
\centering
\includegraphics[width=\textwidth]{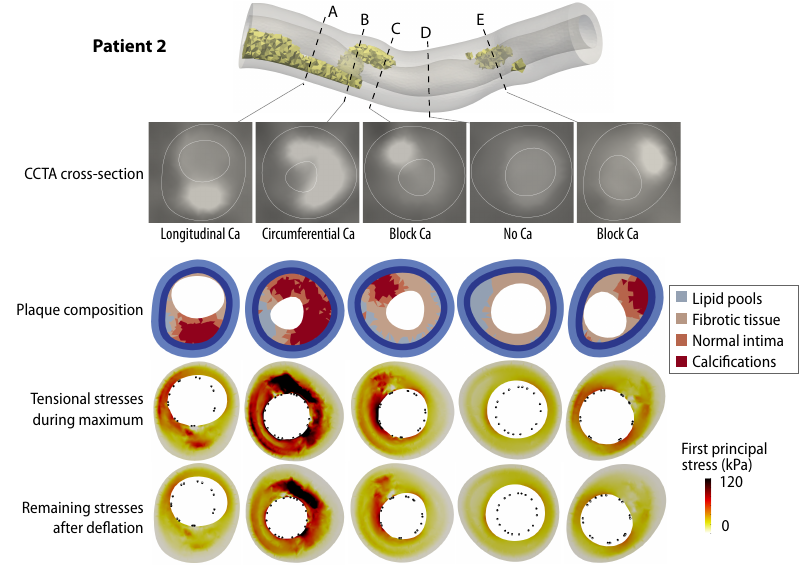}
\caption{First principal stress values in selected cross-sections during and after PCI; Patient 2. The plaque morphology is shown in the \revb{CCTA} cross-sections and in the model. The stent cross-section is shown as black dots. Ca: Calcification. }
\label{fig:pat10_crosssec}
\end{figure}

Patient 2 shows the following plaque characteristics in the cross-sections: Longitudinal or one-sided block calcifications (Slice A, C, E) and circumferential calcification (Slice B). 
Slice D shows no calcifications. 
The highest \revb{tensile} stresses during and after PCI occur in Slice B, where the calcification pattern switches from a longitudinal band to circumferential. 
High stresses in the healthy tissue near the circumferential calcification also affect the media and adventitia. 
In the areas with asymmetric block calcifications, higher stress values occur in the thin-walled areas next to or opposite the calcification.

\clearpage

\begin{figure}[h]
\centering
\includegraphics[width=\textwidth]{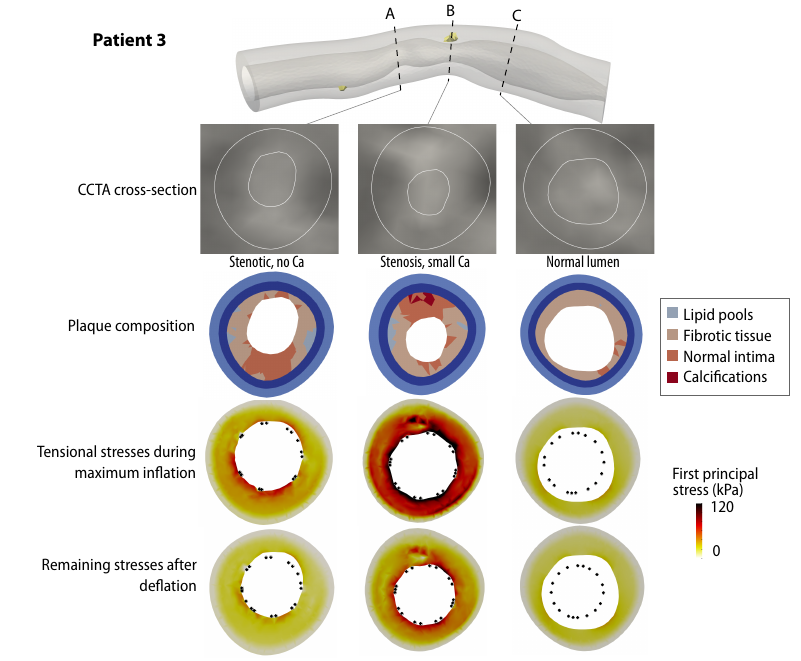}
\caption{First principal stress values in selected cross-sections during and after PCI; Patient 3. The plaque morphology is shown in the \revb{CCTA} cross-sections and in the model. The stent cross-section is shown as black dots. Ca: Calcification. }
\label{fig:pat12_crosssec}
\end{figure}

Patient 3 shows no significant calcifications; the lesion mainly comprises fibrous tissue. 
We observe the highest stresses during the maximum balloon inflation in the location of the most severe stenosis. 
However, the remaining stresses after the balloon withdrawal are reduced drastically and more evenly distributed in the tissue than in the lesions with calcifications. 
\clearpage

\begin{figure}[h]
\centering
\includegraphics[width=\textwidth]{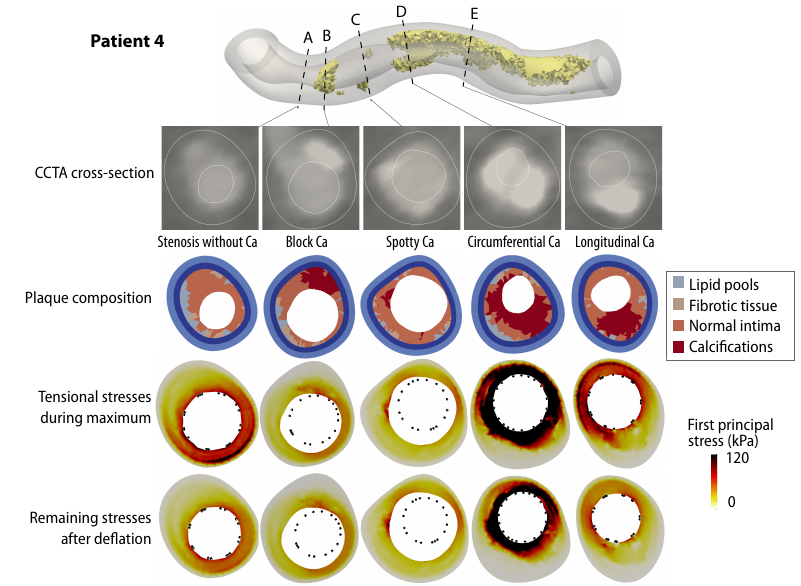}
\caption{First principal stress values in selected cross-sections during and after PCI; Patient 4. The plaque morphology is shown in the \revb{CCTA} cross-sections and in the model. The stent cross-section is shown as black dots. Ca: Calcification. }
\label{fig:pat14_crosssec}
\end{figure}

The lesion in Patient 4 contains a large calcification, that is, similar to the pattern of Patient 2, arranged longitudinally with a switch to a circumferential pattern. 
Additionally, in the proximal section, there are smaller calcification blocks. 
In the proximal part, we observe high stresses opposite of asymmetric calcifications, especially in the thin-walled sections. 
The highest stress values occur in the section of the circumferential calcification. 
The media and adventitia experience lower stresses in calcified areas.


\cleardoublepage

\bibliographystyle{elsarticle-num} 
\bibliography{bibliography3}

\end{document}